\documentclass[a4paper]{article}
\usepackage{amssymb, amsmath}
\usepackage[dvips]{graphicx}
\usepackage{fullpage}
\begin{document}

\title{Correlated Gaussian random walk models of animal dispersal}
\author{Trilochan Bagarti\footnote{Email address: bagarti@iopb.res.in} \\Institute of Physics, Sachivalaya Marg,\\ Bhubaneswar-751005, India}

\maketitle

\begin{abstract}
 A correlated Gaussian random walk(CGRW) model is proposed as a simple model of animal dispersal. The general features of CGRW is described. We will discuss how from this single model a number of different kinds of correlated random walk can be studied. The CGRWs in one dimension is studied in detail and the special limiting cases are discussed where the probability densities are found analytically. Numerical simulations are performed and the results are found to be in good agreement with the theoretical predictions. Directional persistence in CGRWs is discussed for the one dimensional cases. We will show that correlation does not always give rise to directional persistence.
\end{abstract}

\section{Introduction}
Random walk models have been used to study a large number of phenomena in physics, chemistry and biology \cite{chandra, codling}. In biology it has been used to describe various processes occuring at microscopic level inside a cell and also at macroscopic level to understand the motion of the animals in its habitat. The simple random walk model is useful in describing the purely random motion which does not show any correlation. The simplest example would be the diffusion of particles in a media. However, in a large number of cases it has been observed that the random motions in nature are correlated \cite{bart, berg, benha}. The dispersal of animals, birds, insects etc show correlation and persistence. In the correlated random walk(CRW) models the directions of the consecutive steps are correlated. CRWs have been studied extensively since the work of Taylor\cite{taylor}, Goldstein\cite{gold} and Patlak \cite{ patlak}. CRWs on lattices have been studied by a number of authors (see Ref. \cite{gillis, domb, asha, renshw, chen1, chen2, panos}). It has been found that the CRWs are directionally persistent \cite{patlak,  nouv, claes, wu}. The motion of insects like beetels, ants, butterflies etc has been studied theoretically using CRWs by a number of authors \cite{nouv, wallin, byers}. Two dimensional models of CRW has been used to describe the motion of amoeba \cite{hall} and insects \cite{shige}. In these models \cite{hall, shige} consecutive steps are correlated via the relative angle with step size being a random variable. A similar model with constant step size had been studied Concepcion and Argyrakis \cite{concep}. Bovet and Benhamou \cite{bovet}has applied the CRW to explain the spatial patterns of search paths of foraging animals. An interesting work on the postural sway of human body had been modeled Collins and De Luca \cite{collins} using CRW.  Application of CRW models to animal search strategies has been studied Bartumes et. al. \cite{bart}. They propose that some animals may have evolved the ability of performing Levy walks as adaptive strategies to face search uncertainities. 

 Our main aim in this work is to study CRWs which can be applied to the problem of animal dispersal in general. In this article we will study the correlated Gaussian random walk (CGRW) model. In Sec. 2 we shall define the CGRW model in and discuss the general features of the model. We will see how the behavior of the random walk depends on the choice of `strategy' and the `memory' of the random walker. It is found that `strategy' and `memory' both play a very crucial role in animal dispersal. We will also identify how various parameters of our model are related to the `strategy' and `memory' of the walker. Further more, we will show that correlation between consecutive step does not necessarily lead to directional persistence. In Sec. 3 we discuss the symmetric Gaussian random walk(SGRW) in one dimension. The probability densities for the limiting cases are calculated and are compared with the numerical simulations in Sec. 4. In Sec. 5 the directional persistence in these random walks has been discussed. Finally we conclude in Sec. 6 with possible application to correlated motion of animals.
\section{The CGRW in $R^d$}
A Gaussian random walk in $d$ dimension of length $n$ is defined by
\begin{equation}
\mathbf{S}_n = \sum_{k=1}^n\mathbf{x}_k,
\end{equation}
where $\mathbf{S_n} \in R^d$ is the displacement of the random walker from the origin after $n$ steps and $\mathbf{x}_k \in R^d$ is the $k$th step taken by the random walker. If the collection of random vectors $\mathbf{x}_1, \mathbf{x}_2,\ldots \mathbf{x}_n$ are independent and identically  distributed Gaussian random variables then Eq. (1) defines a simple Gaussian random walk(SGRW) in $R^d$. A CRW can be defined in the following manner. The probability for each step $\mathbf{x}_k$  that the random walker takes in the $k$th step depends on a function of the previous step $\mathbf{x}_{k-1}$ denoted by $\mu(\mathbf{x}_{k-1}) \in R^d$ for all $k=2,3,\ldots,n$. Note that $\mu(\mathbf{x}_0)  = \mu_0$ is itself a random vector with a given probability density $p(\mu_0)$ which can be specified by considering the symmetry of the problem. We define a CGRW by Eq. (1) and the Gaussian probability density for the step $\mbox{x}_k$
\begin{equation}
P(\mathbf{x}_k) = \frac{1}{\sqrt{2 \pi}\sigma}\exp\left(\frac{-|\mathbf{x}_k-\mu(\mathbf{x}_{k-1})|^2}{2\sigma^2}\right),
\end{equation}
for all $k=2,3,\ldots,n$. We observe that a more general CRW can be defined with $\mu $  depending not only on the previous step but all the previous steps.
The probability density $P(\mathbf{S}_n, \mu(\mathbf{x}_0))$ is given by
\begin{equation}
P(\mathbf{S}_n, \mu(\mathbf{x}_0)) = \int \delta(\mathbf{S}_n-\sum_i^n\mathbf{x}_i)\prod_{j=1}^n P(\mathbf{x}_j) d\mathbf{x}_j.
\end{equation} 
Note that $\mu(\mathbf{x}_0)$ is the initial condition. In general it is not possible to integrate the right hand side of Eq.(3) however we will see that in one dimension for certain choice of the function $\mu(\mathbf{x})$ we will obtain the probability density $P(\mathbf{S}_n, \mu(\mathbf{x}_0))$ within some approximations. The probability density $P(\mathbf{S}_n) = \int P(\mathbf{S}_n, \mu(\mathbf{x}_0)) p(\mu_0) d\mu_0$ can then be determined from Eq.(3).

The Eqs. (1) and (2) with a given $p(\mu_0)$ completely define our CGRW in $R^d$. Let us now discuss how correlation arises in this random walk. In Fig. (2) we show a schematic of the random walk. The steps $\mathbf{x}_{k-1}$ and $\mathbf{x}_k$ are denoted by solid arrows. Also the function $\mu(\mathbf{x}_{k-1})$ is a vector which is denoted by the arrow with dashed line. The probability that the random walker takes a random step $\mathbf{x}_k$ depends on the length $|\mathbf{x}_{k}-\mu(\mathbf{x}_{k-1})|$ and is given by $P(\mathbf{x}_k) d\mathbf{x}_k$. The expression in Eq. (2) suggests that for a fixed $\sigma$ and a given $\mu(\cdot)$, most of the time the random walker will take step that does not deviate very drastically from its previous step. This gives rise to a correlation in the random walk. Now suppose that we have $\sigma$ sufficiently large such that  $|\mu(\mathbf{x})| \ll \sigma$ for all $\mathbf{x} \in R^d$ then the ratio $|\mathbf{x}_{n}-\mu(\mathbf{x}_{n-1})|/2\sigma^2 \simeq |\mathbf{x}_{n}|/2\sigma^2$ so that CGRW becomes a SGRW having no correlations. In the other hand if $\sigma$ is small then the CGRW will be very highly correlated. We can associate important physical meanings to $\sigma$ and $\mu$. Since the $k$th random step is affected by the $(k-1)$th step through the function $\mu$ it can be called the {\it strategy}  and the quantity $1/\sigma$ acts as a measure of how closely it follows its strategy $\mu$. The quantity $1/\sigma$ is the {\it memory} the random walker possess. 

\begin{figure}[here]
\begin{center}
\includegraphics[width=60mm]{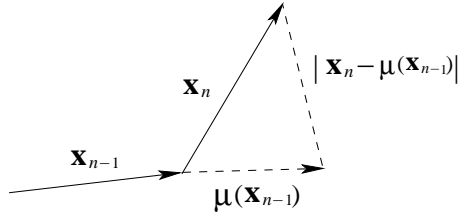}
\caption{Schematic of CGRW in two dimensions.}
\end{center}
\end{figure}

Before we go to the next section let us have a quick look at the the kind of CRWs we shall be studying. In Fig. 1 we have plotted a collection of typical trajectories of CGRWs in $R$ that start from the origin. The length of each random walk is $n=500$ and the Fig. 1(a) and (b) show 100 such trajectories. The trajectories in Fig. 1(a) shows persistence where as (b) does not. We will discuss how the CGRW in Fig. 1(b) has almost `zero directional persistence'. The random walkers here follow two different strategies so it is not surprising that a change of strategy can change the nature of the walks. 

\begin{figure}[here]
\begin{center}
\includegraphics[width=60mm]{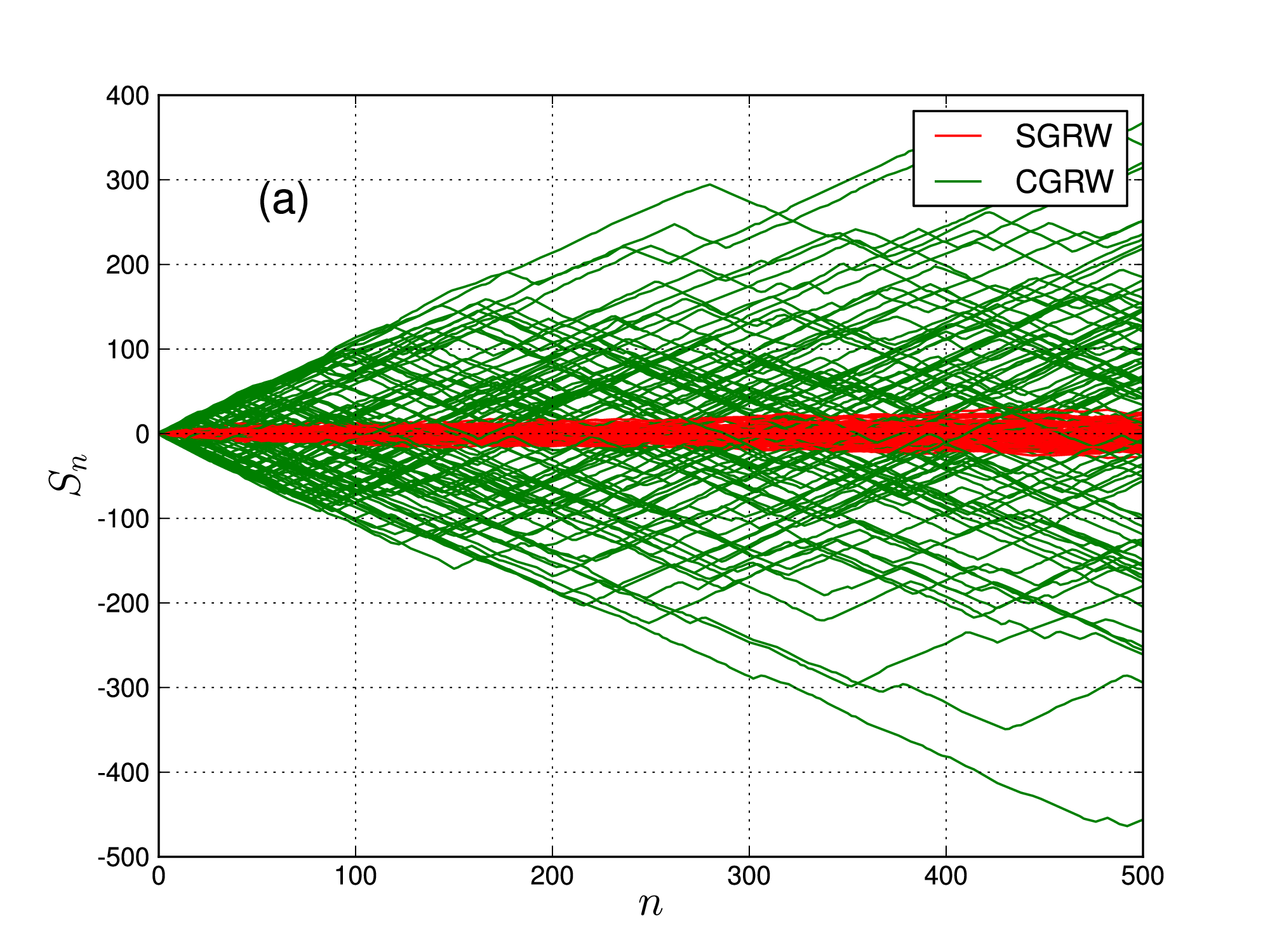}
\includegraphics[width=60mm]{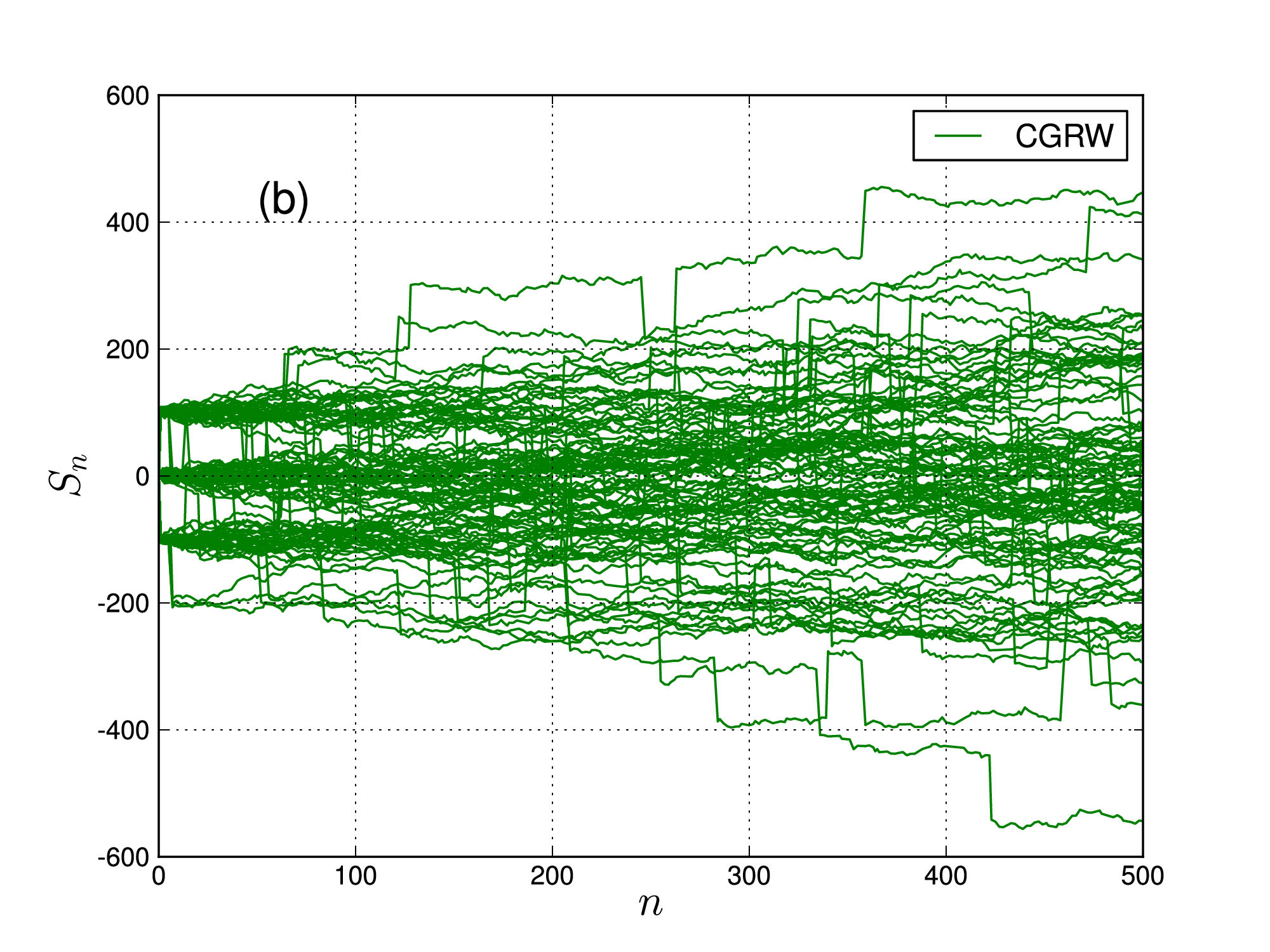}
\caption{Typical trajectories of GRWs of length $n=500$. (a) Trajectories of CGRWs (in green) with correlation depending only on step direction, the trajectories in red are those of SGRWs, the values $\sigma = 0.5$ has been used. (b) Trajectorie of CGRW for $\sigma =2.5$, here correlation depends on both direction and the size of the steps.}
\end{center}
\end{figure}

\section{Symmetric CGRW in $R$}
A symmetric CGRW in one dimension is defined by Eq. (1) and (2) with $d=1$ and $p(\mu_0)$ uniform. We call it a symmetric CGRW because all possible $\mu_0$ can occur with the same probability. Here a uniform $p(\mu)$ implies that the walker will take the first step to the left or to the right with equal probability since $\mbox{sgn}(\mu_0) = \pm 1$ with probability $1/2$. From Eq. (3) the probability density $P(S_n,\mu(x_0))$ for a given $\mu(x_0)$ can be written as
\begin{equation}
P(S_n, \mu(x_0)) = N \int \frac{dk}{2 \pi}\mbox{e}^{-i k S_n}F_n(k,\mu(x_0)),
\end{equation}
where $N=(\sqrt{2 \pi} \sigma)^{-n}$ and the function $F_n(k,\mu(x_0))$ is 
\begin{equation}
F_n(k,\mu(x_0)) = \int \exp\left(\sum_{i=1}^n\frac{-|x_i-\mu(x_{i-1})|^2}{2 \sigma^2}+i k x_i\right)\prod_{j=1}^ndx_j,
\end{equation}
From Eq. (5) we obtain  the following recursion relation
\begin{equation}
F_n(k,\mu(x))=\int\exp\left(\frac{-|y-\mu(x)|^2}{2 \sigma^2}+i k y\right) F_{n-1}(k,\mu(y))dy,
\end{equation}
where $n\geq1$ and $F_0(k,\mu) =1$. We shall use Eqs. (4-6) to obtain the probability density $P(S_n)$ for the cGRWs that we discuss in the following.

The choice of the function $\mu(x)$ decides the properties of the random walks. The simplest case is when $\mu(x) = c$ is a constant. If the $c$ is zero then we have a SGRW which has no correlations between consecutive steps. Also for nonzero $c$ we shall again have a SGRW which is baised in one direction depending on the sign of the constant $c$. An interesting long tailed behavior of the probability density can arise if $\mu(x)$ is diverging at some values of $x$. A choice of $\mu(x) = 1/x$ can give rise to such long tailed behavior. There could be a number of CGRWs which may be interesting but most of them would be too difficult to understand analytically. Here we discuss two different random walks for which we can calculate approximate analytical results that gives us insight into the nature of these correlated random walks.  First we will discuss a CGRW in which every step that the random walker takes depends only on the direction of the previous step. The second type of CGRW that we discuss will not only depends on the direction of the previous step but also on the size of the step (see Fig.1).

\subsection{CGRW  with correlation depending on direction of the steps}
We define $\mu(x_k) = \hat{x}_k$ where $\hat{x}_k = \mbox{sgn}(x_k)$ and $\mu_0=\pm1$ with probability $1/2$. With this setting  Eq. (1) and (2) in one dimension defines a symmetric CGRW. Substituting $\mu(x)=\hat{x}$ in Eq. (6) we obtain a recursion relation for $F_n(k,\hat{x})$ which can be written in the following matrix form
\begin{equation}
\mathbf{F}_n = \sqrt{\frac{\pi}{2}}\sigma e^{-k^2 \sigma^2/2}\mathbf{M}\mathbf{F}_{n-1},
\end{equation}
where $\mathbf{F}_n =(F_n(k,-1),F_n(k,1))^T$ and $\mathbf{M}$ is a $2\times2$ matrix given by
\begin{equation}
\mathbf{M} = \left(\begin{array}{cc}
                     e^{-ik}(1-\mbox{erf}(\frac{-1+ik\sigma^2}{\sqrt{2}\sigma}))  & e^{-ik}(1+\mbox{erf}(\frac{-1+ik\sigma^2}{\sqrt{2}\sigma})) \\
                     e^{ik}(1-\mbox{erf}(\frac{1+ik\sigma^2}{\sqrt{2}\sigma})) & e^{ik}(1+\mbox{erf}(\frac{1+ik\sigma^2}{\sqrt{2}\sigma})) 
                    \end{array} \right).
\end{equation}
Once $F_n(k,\hat{x})$ is obtained from Eq. (7) we can find the density $P(S_n,\hat{x}_0)$ from Eq. (4).
Note that if $x,y \in R$ then $\mbox{Re}(\mbox{erf}(x+iy)) = -\mbox{Re}(\mbox{erf}(-x+iy))$ and $\mbox{Im}(\mbox{erf}(x+iy)) = \mbox{Im}(\mbox{erf}(-x+iy))$. Setting $x=1/\sqrt{2}\sigma$, $y=k\sigma/\sqrt{2}$ and defining $A=\mbox{Re}(\mbox{erf}(x+iy))$ and $B=\mbox{Im}(\mbox{erf}(x+iy))$, the eigenvalues of $\mathbf{M}$ are
\begin{equation}
\epsilon_{1,2}=(1+A)\cos k -B \sin k\pm\left[-4A+((1+A)\cos k-B \sin k)^2\right]^{\frac{1}{2}}.
\end{equation}
From the recursion relation Eq. (7) we have 
\begin{equation}
\mathbf{F}_n = \left( \frac{\pi}{2}\right)^{n/2}\sigma^n e^{-nk^2\sigma^2/2}\mathbf{M}^n \mathbf{F}_0.
\end{equation}
The matrix $\mathbf{M}^n$ can be written in terms of the eigenvalues $\epsilon_{1,2}$ and $\mathbf{M}$ as 
\begin{equation}
\mathbf{M}^n=\epsilon_1^n\left( \frac{\mathbf{M}-\epsilon_2 \mathbf{I}}{\epsilon_1-\epsilon_2}\right)+\epsilon_2^n\left( \frac{\mathbf{M}-\epsilon_1 \mathbf{I}}{\epsilon_2-\epsilon_1}\right),
\end{equation}
where $\mathbf{I}$ is the $2\times2$ identity matrix. Substituting Eq.(11) in Eq.(10) and using Eq. (4) we obtain
\begin{eqnarray}
\mathbf{Q}_n &=& 2^{-n} \int \frac{dk}{2 \pi}e^{-nk^2\sigma^2/2-i k S_n}\\ \nonumber
&&\times \left[ \epsilon_1^n\left( \frac{\mathbf{M}-\epsilon_2 \mathbf{I}}{\epsilon_1-\epsilon_2}\right)+\epsilon_2^n\left( \frac{\mathbf{M}-\epsilon_1 \mathbf{I}}{\epsilon_2-\epsilon_1}\right)  \right]\mathbf{F}_0,
\end{eqnarray}
where $\mathbf{Q}_n =(P(S_n, -1), P(S_n,1))^T$ and since $F_0(k,\pm1)=1$ with probability $1/2$ the probability density $P_{\mbox{\scriptsize cGRW}}(S_n) =(P(S_n, -1)+P(S_n,1))/2 $ can be obtained from Eq. (12). Although it is difficult to obtain a closed form expression for the integral in the right hand side of Eq. (12), it can always be integrated numeically for any values of $\sigma$ and can be compared with the probability density obtained from numerical simulations. In Sec. 4 we will use the expression in Eq. (12) to obtain probability densities for the limiting cases $\sigma \ll 1$ and $\sigma \gg 1$.

\subsection{CGRW with correlation depending on both direction and magnitude of the steps}
Now we present the second type of random walk in which $\mu(x)$ depends also on the magnitude of x. Let us define $\mu(x)$ by the function
\begin{equation}
\mu(x) = \left\{ \begin{array}{rr}
                        \mbox{sgn}(x) p, ~~~\mbox{if~~} |x|\le x_0, \\
                        0, ~~~~~~~~\mbox{otherwise},
                     \end{array}
             \right.
\end{equation}
where $p >0$. The CGRW for the above choice of $\mu(x)$ has completely different properties from the one we discussed in Sec. 3.1. Substituting the above $\mu(x)$ in Eq. (6) we obtain the same matrix recursion relation Eq. (7) with $\mathbf{F}_n=(F_n(k,-p), F_n(k,0), F_n(k,p))^T$ and  $\mathbf{M}$ is now a $3 \times 3$ matrices
\begin{equation}
\mathbf{M}=\left( \begin{array}{lll}
               e^{-i p k}\psi_1(-p) & e^{-i p k}\psi_2(-p) & e^{-i p k}\psi_3(-p) \\
               \psi_1(0) & \psi_2(0) & \psi_3(0) \\
               e^{i p k}\psi_1(p) & e^{i p k}\psi_2(p) & e^{i p k}\psi_3(p) \\
            \end{array}
     \right)
\end{equation}
where $\psi_1(\mu) = \mbox{erf}(\frac{\mu+i k \sigma^2+x_0}{\sqrt{2}\sigma})-\mbox{erf}(\frac{\mu+i k \sigma^2}{\sqrt{2}\sigma})$, $\psi_2(\mu)=2-\mbox{erf}(\frac{\mu+i k \sigma^2+x_0}{\sqrt{2}\sigma})+\mbox{erf}(\frac{\mu+i k \sigma^2-x_0}{\sqrt{2}\sigma})$ and  $\psi_3(\mu) = \mbox{erf}(\frac{\mu+i k \sigma^2}{\sqrt{2}\sigma})-\mbox{erf}(\frac{\mu+i k \sigma^2-x_0}{\sqrt{2}\sigma})$. An expression of the form Eq.(12) can be obtained for this case by diagonalizing $\mathbf{M}$ but will be too cumbersome to write it here. Such an expression can however always be numerically integrated to compare with the results obtained from numerical sumulations. We will use the recursion relation Eq.(7) with the matrix in Eq. (14) while discussing the limiting cases in the following section. We will find an approximate solution to the recursion relations and we will see that the results agree quite well with the numerical simulations.

\section{Approximate solutions and numerical results}
It is in general difficult to find a closed form expression for $P(S_n)$ using Eq.(12). Here we discuss the approximate solutions when $\sigma \ll 1$ and $\sigma \gg 1$. When $\sigma \gg 1$ we can approximate $|\mathbf{x}_k-\mu(\mathbf{x}_{k-1})|^2/2\sigma^2 \simeq |\mathbf{x}_k|^2/2\sigma^2$ and the CGRW approaches SGRW. The probability density is $P_{\mbox{\scriptsize SGRW}}(S_n)=\exp(-S_n^2/2 n \sigma^2)/\sqrt{2n \pi} \sigma$ . On the other hand when $\sigma \ll 1$, $P(S_n)$ for the CGRW shows a bimodal behavior which can be written as the sum of two Gaussians centered around $n$ and $-n$.  Note that in the limit $k\rightarrow \infty$ and $\sigma>0$, $\mbox{erf}((\pm1+ik\sigma^2)/\sqrt{2}\sigma) e^{-k^2\sigma^2/2} =0$ and for $k\rightarrow 0$ it is $\mbox{erf}(\pm1/\sqrt{2}\sigma)$. The significant contribution to the integral in Eq. (12) will come from a small neighborhood of $k=0$. Expanding $\mbox{erf}((\pm1+ik \sigma)/\sqrt{2}\sigma)$ in Taylor series near the origin we have 
\begin{equation}
\mbox{erf}((\pm1+ik \sigma)/\sqrt{2}\sigma)=\mbox{erf}(\pm1/\sqrt{2}\sigma)+\lambda(k^2+i2k)+O(k^3),
\end{equation}
where $\lambda=\sigma e^{-1/2\sigma^2}/\sqrt{2\pi}$. Further for $\lambda \ll \sigma \ll 1$ we take the limit $\lambda \rightarrow 0$ and we can write from Eq. (15), $A=\pm1$ and $B=0$. Substituting these values in Eq. (9) we have $\epsilon_1 = 2e^{-ik}$ and $\epsilon_2 = 2e^{ik}$. Now using Eq.(12) the probability density can be written as
\begin{equation}
P_{\mbox{\scriptsize approx.}}(S_n)=\int\frac{dk}{2\pi}e^{-ikS_n-nk^2\sigma^2/2}\cos nk
\end{equation}
The probability density $P(S_n)$ for the limiting cases is given by
\begin{equation}
P_{\mbox{\scriptsize app.}}(S_n)= \left\{ \begin{array}{ll}
                    \frac{1}{\sqrt{2\pi n} \sigma}\exp(-S_n^2/2 n \sigma^2), \mbox{\hspace{4cm} if } \sigma \gg 1, \\
                    \frac{1}{2\sqrt{2 \pi n}\sigma}\left[ \exp\left( \frac{-(S_n-n)^2}{2n \sigma^2}\right) + \exp\left( \frac{-(S_n+n)^2}{2n \sigma^2}\right)  \right], \mbox{~~~~~~ if } \sigma\ll1.
                   \end{array}
            \right.
\end{equation}
In Fig.(3) we compare the limiting solutions Eq.(17) and the exact results obtained from Eq. (12) with the histograms obtained from numerical simulations.
\begin{figure}[here]
\begin{center}
\includegraphics[width=60mm]{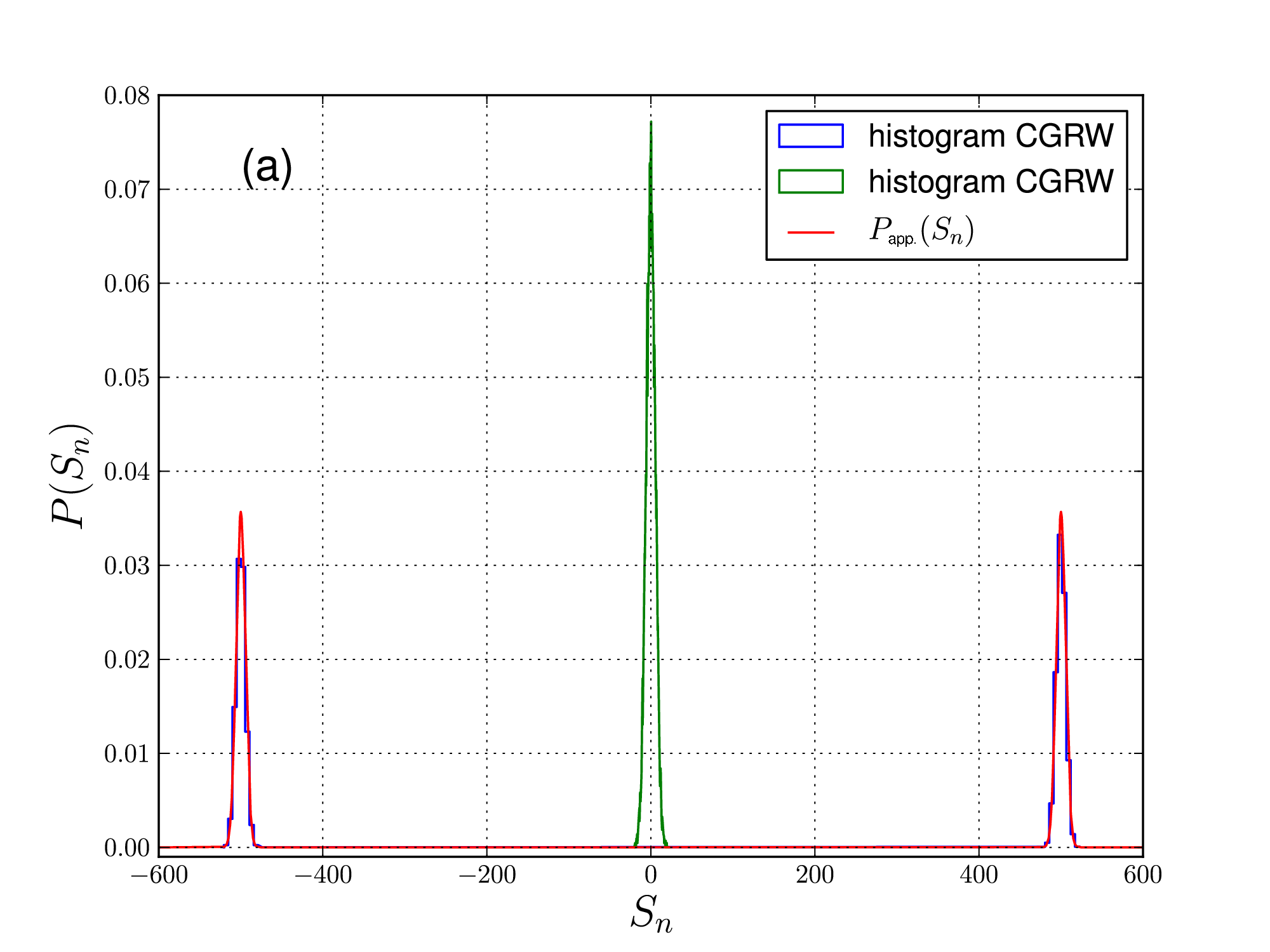}
\includegraphics[width=60mm]{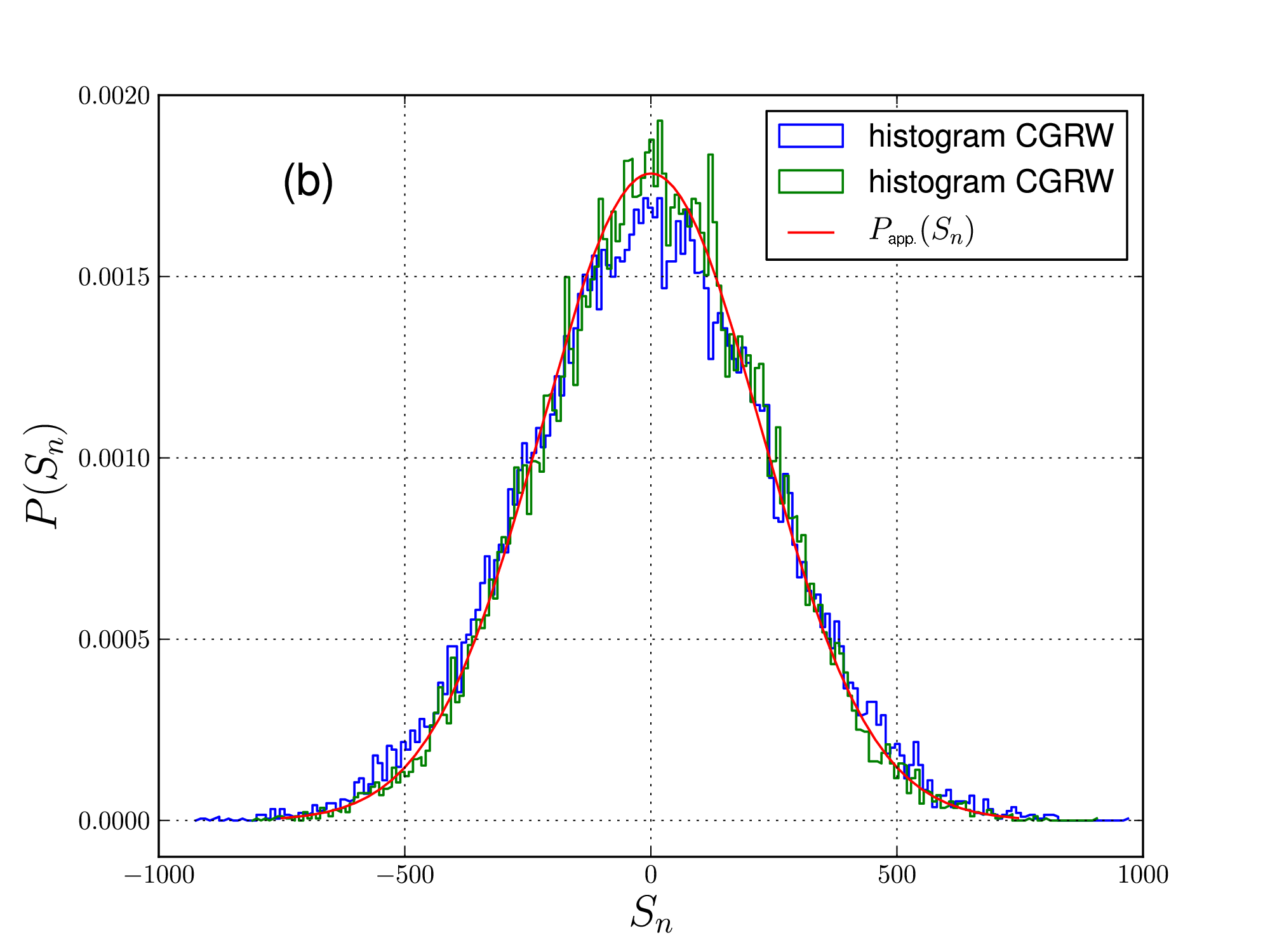}
\includegraphics[width=60mm]{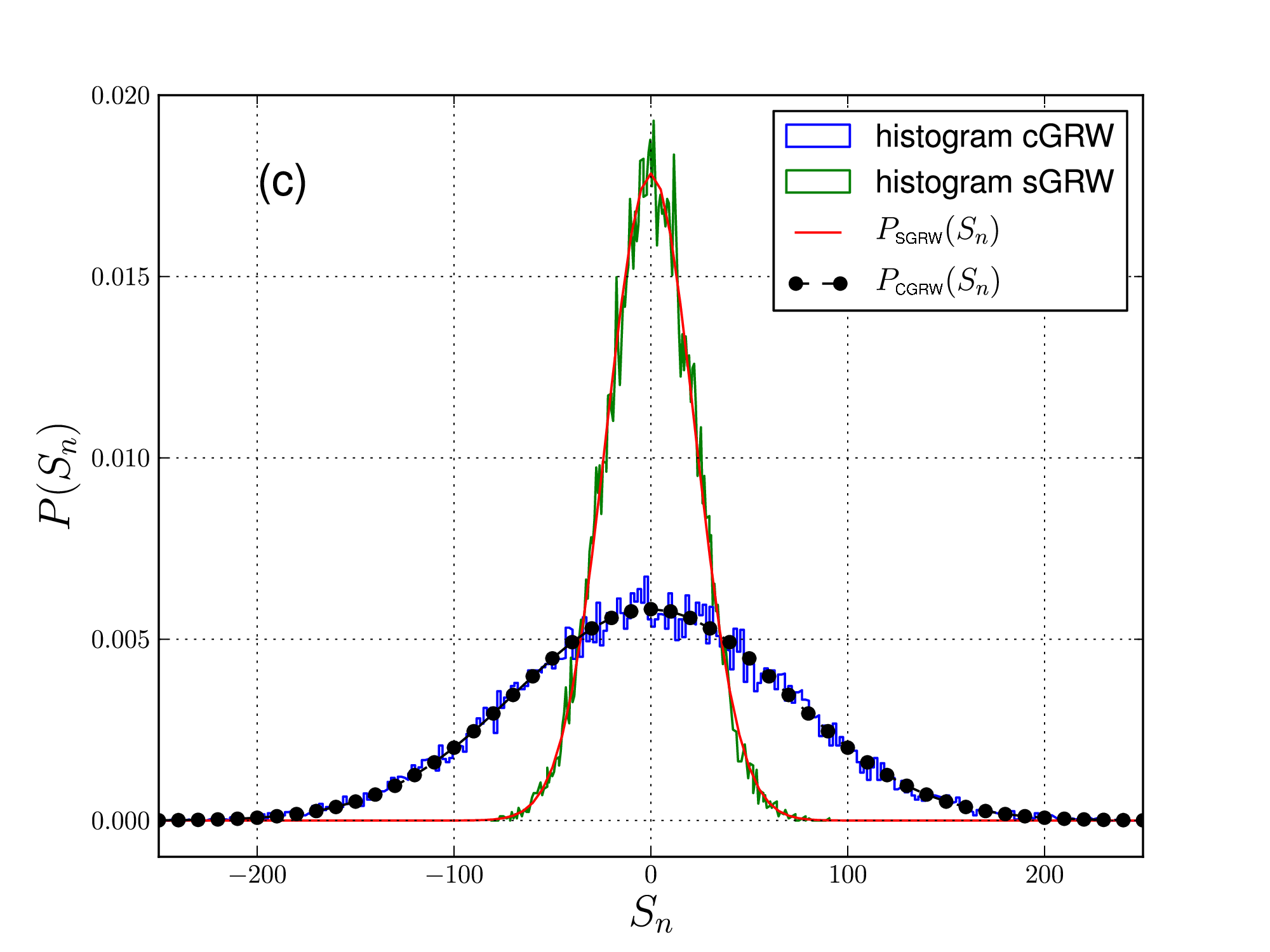}
\includegraphics[width=60mm]{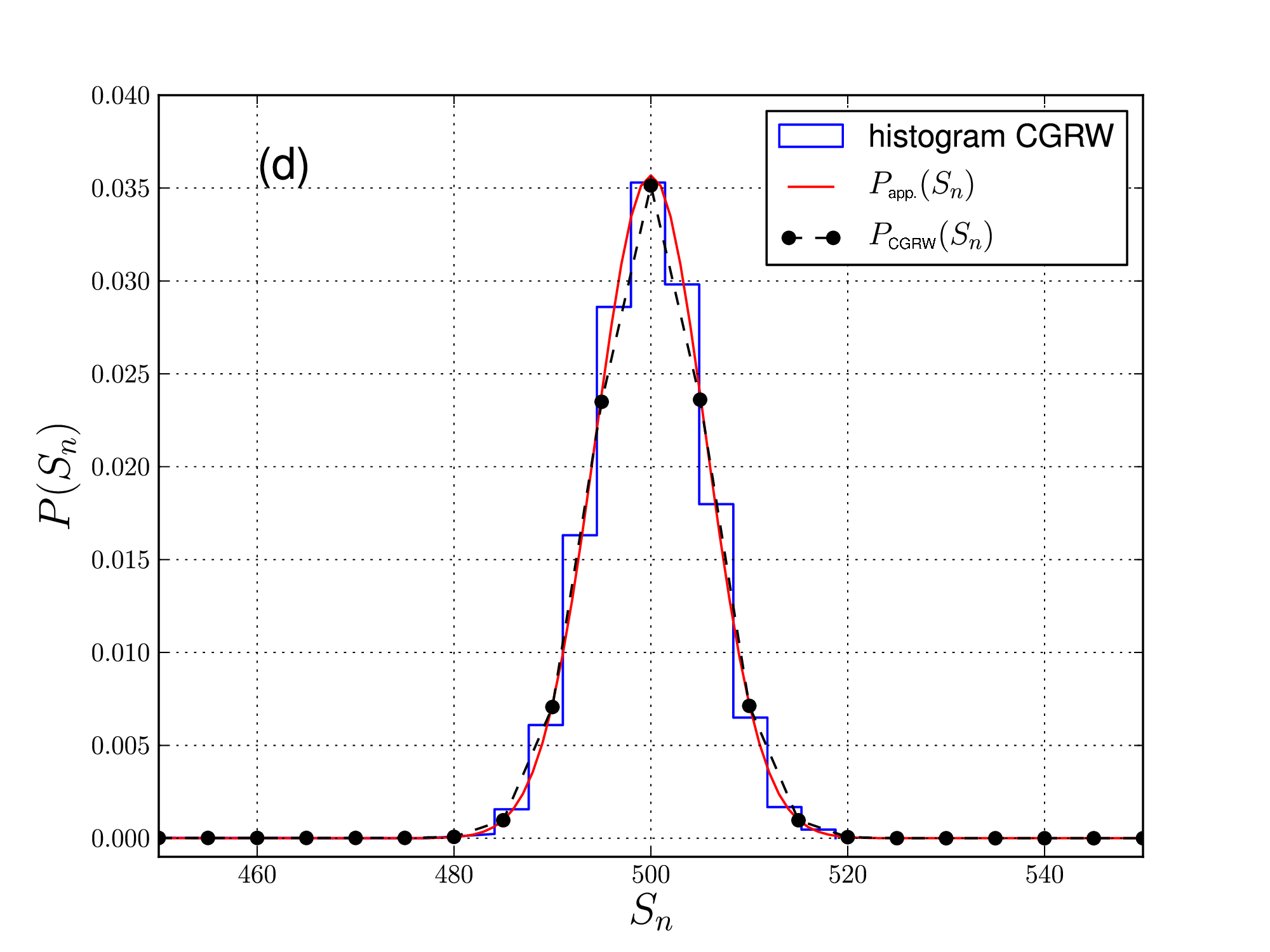}
\end{center}
\caption{A comparison of histograms and probability densities for various cases.  (a) Probability densities for $\sigma=0.25$ shows a bimodality. (b) At $\sigma =10$, CGRW $\rightarrow$ SCGRW. (c) For $\sigma=1$ with the red curve showing $P_{\mbox{\scriptsize SGRW}}(S_n)$.  (d) A comparison of numerically obtained probability density with the exact expression $P_{\mbox{\scriptsize CGRW}}(S)$ from Eq.(12) and the approximate probability density $P_{\mbox{\scriptsize app.}}(S_n)$.}
\end{figure}

We turn next to the CGRW described in Sec. 3.2. In the limiting case $x_0/\sigma \ll 1$, $p/\sigma \gg 1$, the matrix $M$ in Eq. (14) assumes a simple form with $\psi_2(0) \simeq 2[1-\mbox{erf}(\frac{x_0}{\sqrt{2} \sigma})]$, $\psi_1(0) = \psi_3(0) \simeq \sqrt{\frac{2}{\pi}} \frac{x_0}{\sigma} \exp(k^2 \sigma^2/2)$, $\psi_2(\pm p)\simeq 2$ and $\psi_1(\pm p)=\psi_3(\pm p)\simeq 0$. With these approximations we can write from Eq. (7), 

\begin{eqnarray}
F_n(k, \pm p) = \sqrt{2 \pi} \sigma e^{\pm i k p-k^2\sigma^2/2} F_{n-1}(k,0), \\
F_n(k,0)=\sqrt{2 \pi} \sigma e^{-k^2\sigma^2/2} (1-\mbox{erf}(\frac{x_0}{\sqrt{2} \sigma})) F_{n-1}(k,0)+x_0 [F_{n-1}(k, -p)+F_{n-1}(k, p)].
\end{eqnarray}
It is straight forward to solve these recurrence relations but one obtains cumbersome expressions since the recursion relation for $F_n(k,0)$ is second order. In our analysis we will find an approximate solution to the Eq. (18-19). In the limit $x_0 \rightarrow 0$, Eq. (19) becomes independent of $F_{n-1}(k,\pm p)$ and Eq. (18) becomes $F_n(k, \pm p) = e^{\pm i kp} F_n(k,0)$. An approximate solution of Eqs.(18) and (19) can be obtained if we assume $F_{n-1}(k, \pm p) \simeq e^{\pm i kp} F_{n-1}(k,0)$ for $x_0 \neq 0$. We have 
\begin{equation}
F_n(k,0) = \left[\sqrt{2 \pi} \sigma (1-\mbox{erf}(\frac{x_0}{\sqrt{2} \sigma})) \mbox{e}^{-k^2\sigma^2/2}+2x_0 \cos kp \right]^n
\end{equation}
The probability density $P_{\mbox{\scriptsize app.}}(S_n)$ can be written as
\begin{equation}
P_{\mbox{\scriptsize app.}}(S_n) = \int \frac{dk}{6 \pi} \exp(-ikS_n) \left[ \alpha~\mbox{e}^{-k^2 \sigma^2 /2} +\beta \cos kp \right]^n(1+2 \cos kp).
\end{equation}
where $\alpha = (1-\mbox{erf}(\frac{x_0}{\sqrt{2} \sigma}))$ and $\beta = \frac{2x_0}{\sqrt{2 \pi} \sigma}$. In Eq.(21) a factor $1/3$ is multiplyed since $\mbox{Prob}(\mu_0 \in \{-p, 0, p \})=1/3$. 
As in the previous case for large $\sigma$, cGRW approaches the sGRW we see the same behavior here as well, hence $P_{\mbox{\scriptsize app.}}(S_n) = P_{\mbox{\scriptsize SGRW}}(S_n)$. We are rather interested in the other limit which shows interesting oscillating features in the probability density (e.g. see Fig. 4 (a)). Integrating Eq. (21) gives
\begin{equation}
P_{\mbox{\scriptsize app.}}(S_n) = \frac{1}{3}\sum_{r=0}^n \binom{n}{r} \frac{\alpha^{n-r} \beta^r}{2^r \sqrt{2 \pi (n-r)} \sigma} \left[ \sum_{t=0}^r \binom{r}{t}\mbox{e}^{\frac{-(S_n-(r-2t)p)^2}{2(n-r)\sigma^2}}+ \sum_{t'=0}^{r+1} \binom{r+1}{t'} \mbox{e}^{\frac{-(S_n-(r+1-2t')p)^2}{2(n-r)\sigma^2}}  \right].
\end{equation}
In Fig. 4 (a) and (b) we can see Eq. (22) has excelent agreement with the numerical results. In the next section we shall discuss the directional presistence of the two CGRWs we have discussed above. The numerical simulations of the random walks were performed for $n=500$ for values of $0.25 \le\sigma \le 10$. In Figure 3(c), 3(d) $P_{\mbox{\scriptsize CGRW}}(S_n)$ was calculated numerically using the trapezoidal rule from the integral in Eq. (12). Clearly for large $\sigma$ Fig. 3 (b) and 4 (d) shows that the correlations are lost and the cGRW approaches sGRW. The approximate results obtained in Eq.(17) for $\sigma \ll 1$ and Eq.(22) agree very well with our numerical simulations. Fig. 3 (a) shows the bimodal probability density $P_{\mbox{\scriptsize app.}}(S_n)$ at $\sigma =0.25$ and in Fig. 4 (a) $P_{\mbox{\scriptsize app.}}(S_n)$ shows oscillatory pattern where $\sigma =0.5, x_0=0.001$ and $p=100$. The bimodality is a direct consequence of the persistence in random walk. The root mean square displacement is $\sim n$ which suggests that the motion is superdiffusive. In Fig. 4 (a) the oscillatory nature of the probability density tells us that the random walk in the second case consists of clusters of walks between long jumps. This type of walk is actually observed in nature. Imagine for example certain insects, say butterflies. The random motion of a butterfly which when flying over a tree looking for nectar in flowers will consist of hovering over bunch of flowers and a less frequent long flights of a typical. This typical length in our model is $p$  and peaks in Fig. 4(a) denotes the clusters of walks. The CGRW model in Sec. 3.2 describe the random motion of such a butterfly with only exception that the butterfly is myopic. We note that in the real case the butterfy is not myopic and therefore the strategy $\mu$ has to be designed accordingly.

\begin{figure}[here]
\begin{center}
\includegraphics[width=60mm]{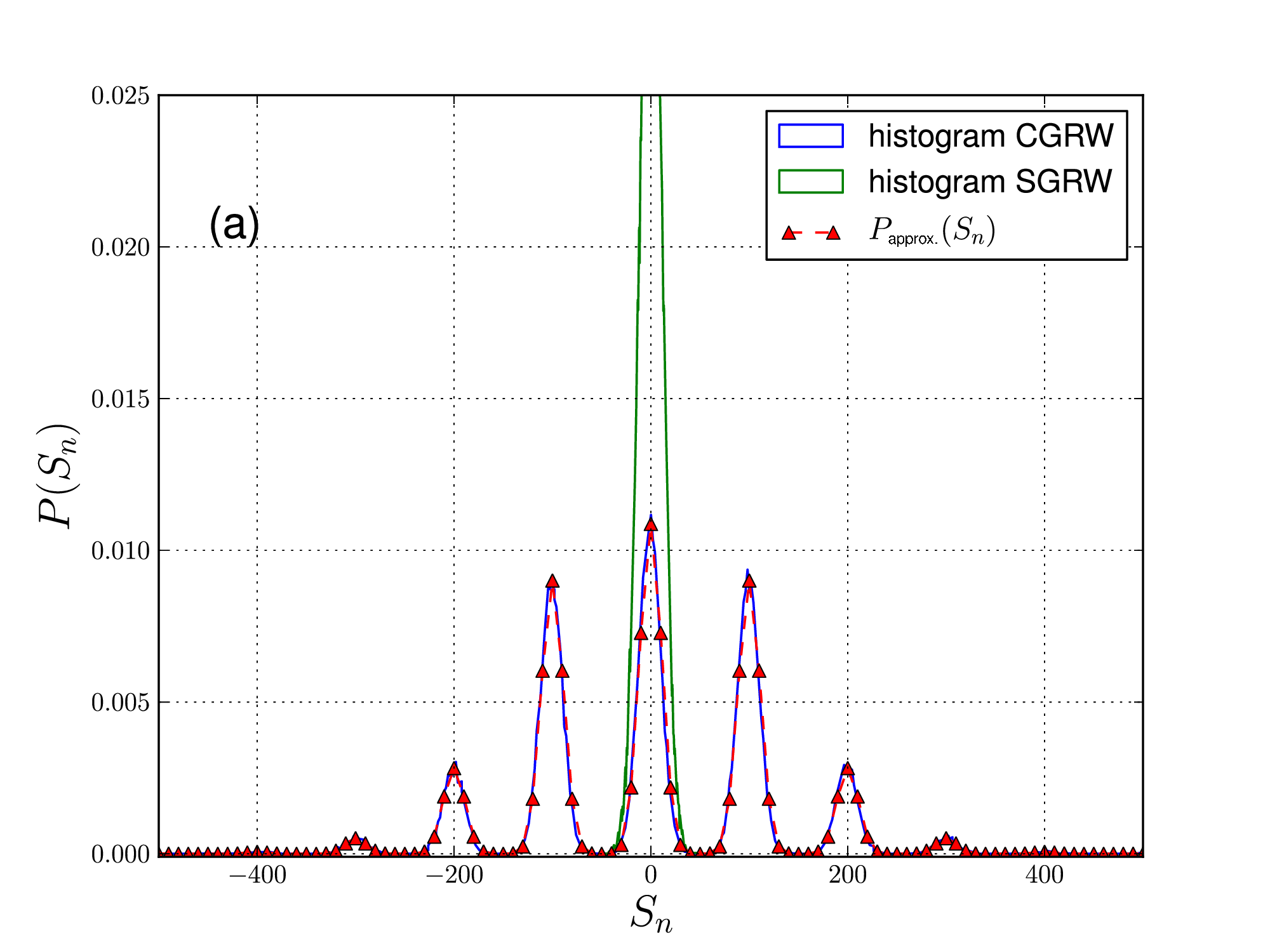}
\includegraphics[width=60mm]{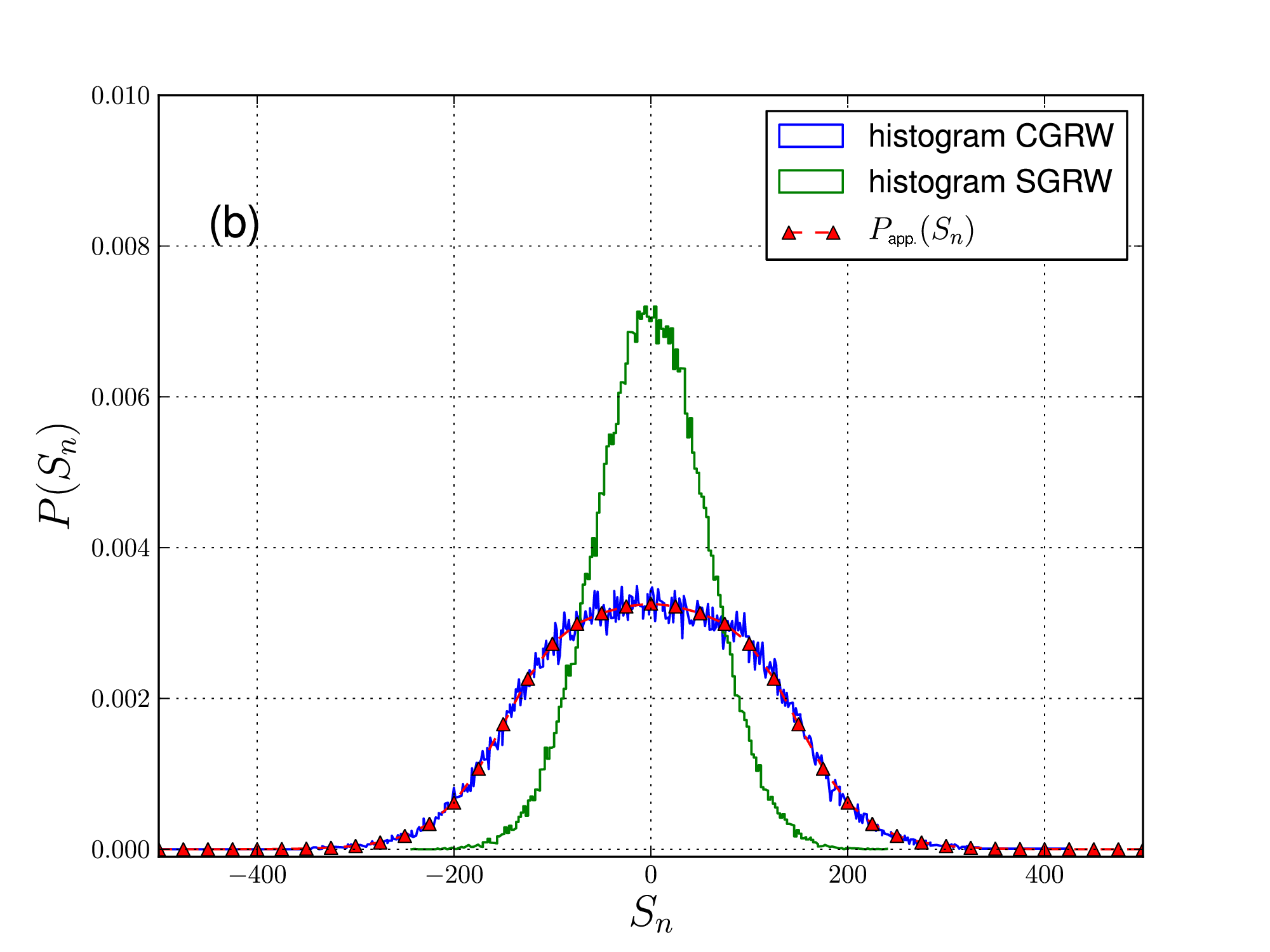}
\includegraphics[width=60mm]{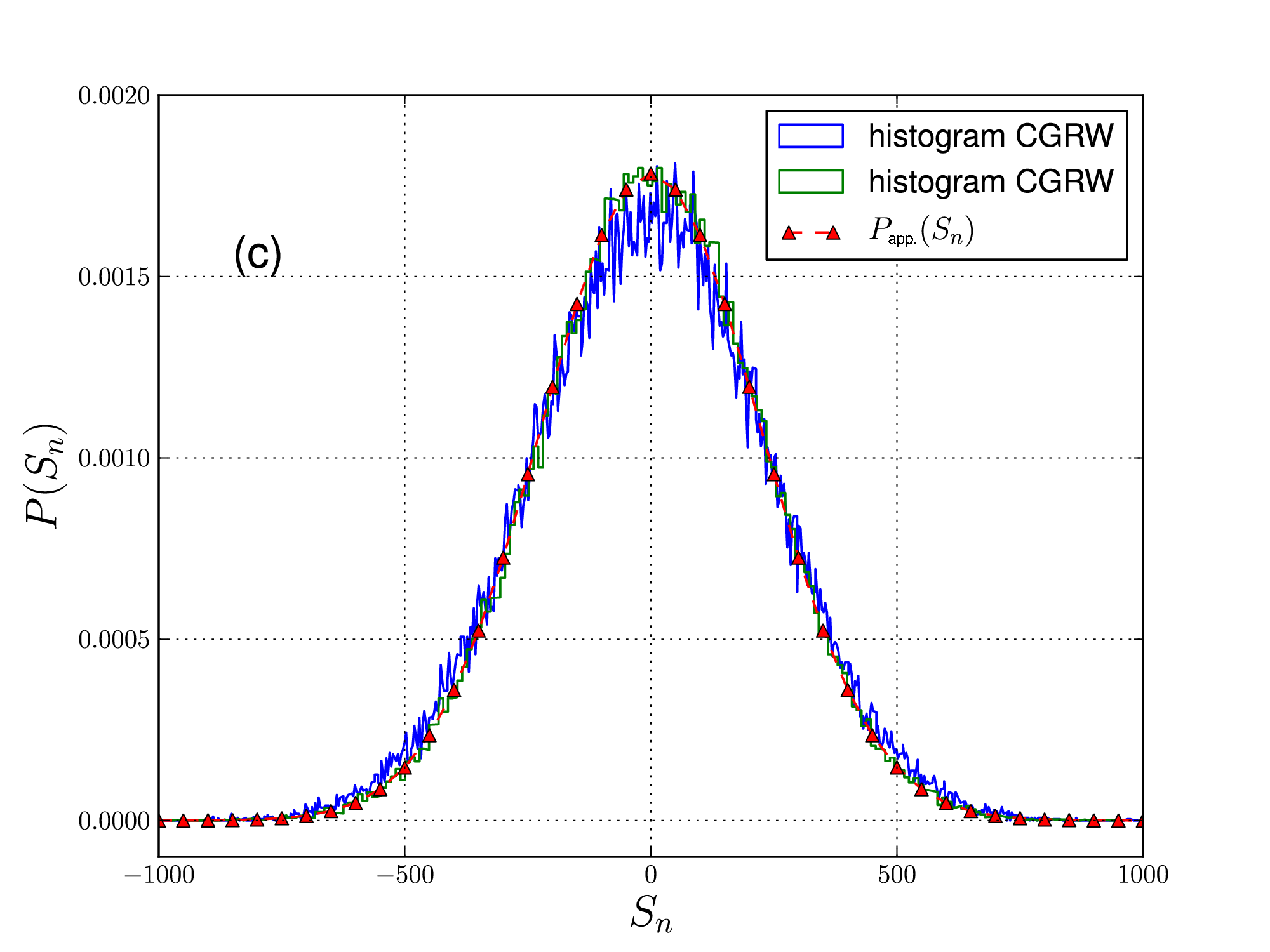}
\includegraphics[width=60mm]{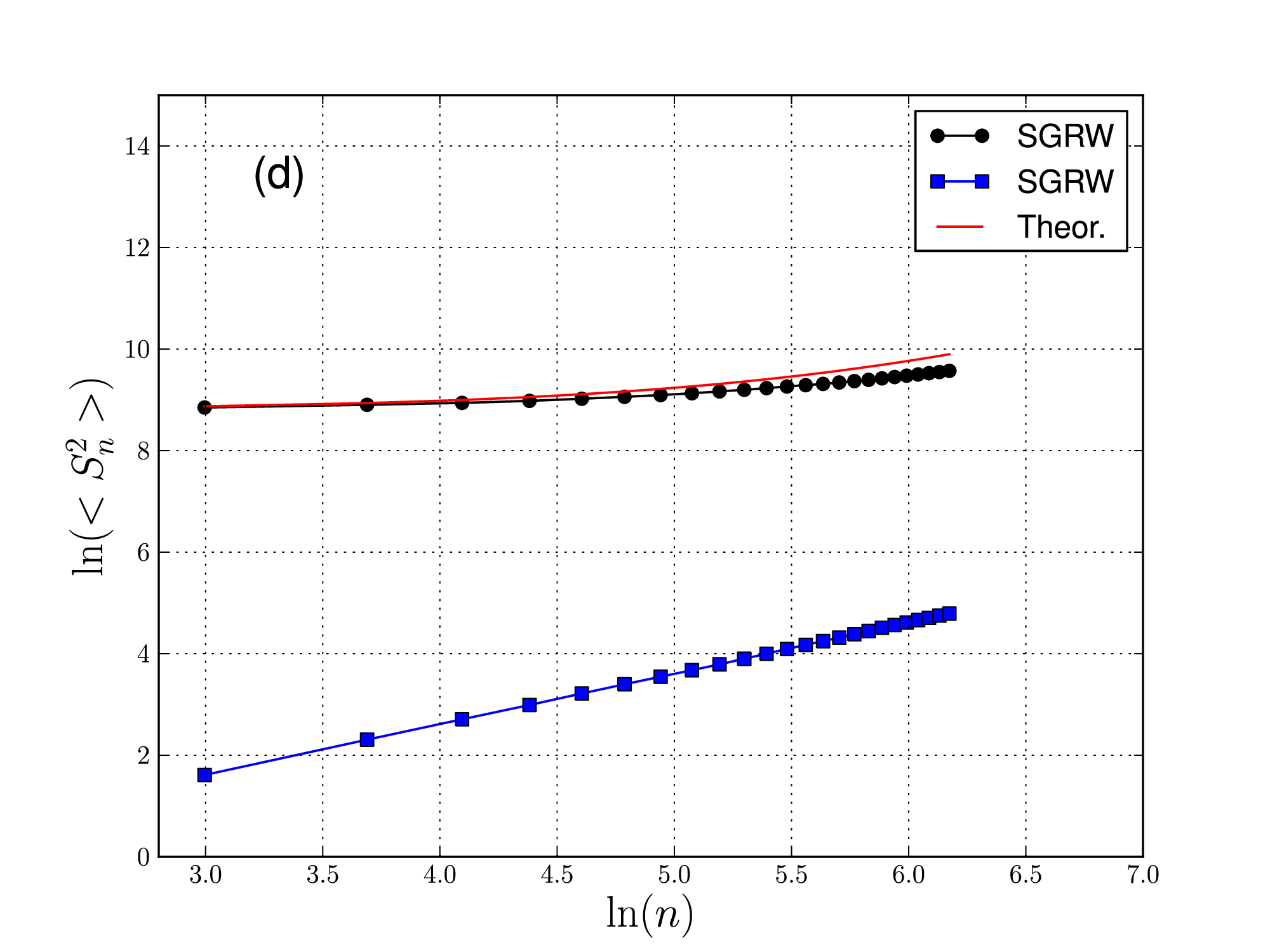}
\end{center}
\caption{A comparison of histograms and probability density $P_{\mbox{\scriptsize app.}}(S_n)$ from Eq.(22)  with $x_0 =0.001$. (a) Probability densities for $\sigma=0.5$ shows an oscillatory behavior, (b) at $\sigma = 2.5$ the probability density has no oscillation and (c) for $\sigma =10.0$, CGRW $\rightarrow$ SGRW, here $P_{\mbox{\scriptsize app.}}(S_n) = P_{\mbox{\scriptsize SGRW}}(S_n)$ (in red). (d) $\ln \langle S_n^2\rangle$  (in red) show the approximate theoretical estimate obtained from Eq. (22) with $x_0=0.001$ and $\sigma = 0.5$, compared with the numerical results, $\langle S_n^2\rangle$ for SGRW is shown in (blue).}
\end{figure}

\section{Persistence in CGRW}
Persistence in the motion is an important property of  CRWs. Directional persistence is the property, by virtue of which, an object in motion prefers to move in the same direction. For the case of a SGRW, since steps are uncorrelated it is very unlikely that the random walker will continue to move in the same direction for a long time. The probability that the random walker will continue to move in the same direction for $n$ consecutive steps is $2^{-n}$. For the CGRW defined in Sec. 3.1 this probability is $p_n=[\frac{1}{2}(1+\mbox{erf}(\frac{1}{\sqrt{2}\sigma}))]^n$. Although these probabilities are monotonically decreasing with the number of steps $n$ for both SGRW and CGRW, we note that $p_n>2^{-n}$ for all $n>0$ and $\sigma$ finite. This suggests that the rate of decrease of $p_n$ is slower than that of $2^{-n}$. It is this property of the CRWs which is called directional persistence. Persistence in random walks can result in amomalous diffusion. From Eq. (17) the mean square displacement $\langle S_n^2 \rangle = n \sigma^2$ for $\sigma \gg 1$ and $\langle S_n^2 \rangle = n (n+\sigma^2)$ for $\sigma \ll 1$ which shows anomalous behavior for $\sigma \ll 1$. Interestingly in the limit $\sigma \rightarrow \infty$ the probability $p_n \rightarrow 2^{-n}$ (see Fig. 5) the  correlations are lost and we say the SGRW has  `{\it zero persistence}'. A random walk with $p_n\ge 2^{-n}$ is a persistent random walk. Recall that we have associated the meaning `{\it memory}' to the quantity $1/\sigma$, hence a vanishing `{\it memory}' leads to a complete loss of persistence.

The CGRW defined in Sec. 3.2 has a different story to say. From Eq. (22) the mean square displacement is found to be $\langle S_n^2 \rangle =(\alpha+\beta)^{n-1}[p^2(2\alpha+(2+3n)\beta)+3n\alpha \sigma^2]/3$. This is an approximate estimate valid only when $x_0/\sigma \ll 1$ and $p/\sigma \gg 1 $ (see Fig. 4(d)). For a vanishingly small $x_0$ but with a finite memory (i. e. $\sigma < \infty$) we have $\langle S_n^2 \rangle \rightarrow (2p^2+3 n \sigma^2)/3$. For any $x_0$ which can be set as close to zero as we please, we will always have steps of length $\sim p$ in the $k$th step when ever the $(k-1)$th step has length less than $x_0$. Now consider any two such consecutive steps of length $\sim p$ at the $k$th and $k'$th time step where $k' > k+1> 0$. The steps will have length $|x_j| \geq x_0$ for all $j=k+1, \ldots, k'-1$. This portion of the CGRW is an unbaised random walk as $\mu(x_j) =0$ for all $j=k+1, \ldots, k'-1$. We also observe that $\mu(x_j) =\pm p$ with probability $1/2$ for $j=k \mbox{  and  } k'$ as the $(k-1)$th and $(k'-1)$th steps can be positive or negative with equal probability. The probability $p_n$ that this CGRW will move in the same direction for $n$ steps will be $2^{-(n-2)}\times 1/2^2 = 2^{-n}$. The CGRW described in Sec. 3.2 therefore has zero persistence. This fact is also true for any finite $x_0$ with $p/\sigma \gg 1$. This suggests that correlation between consecutive steps alone does not always lead to directional persistence in random walks. Further, the calculation of the probability $p_n$ for the real cases of animal dispersal should give a hint of how to design the function $\mu$ and the constant $\sigma$.

\begin{figure}[here]
\begin{center}
\includegraphics[width=60mm]{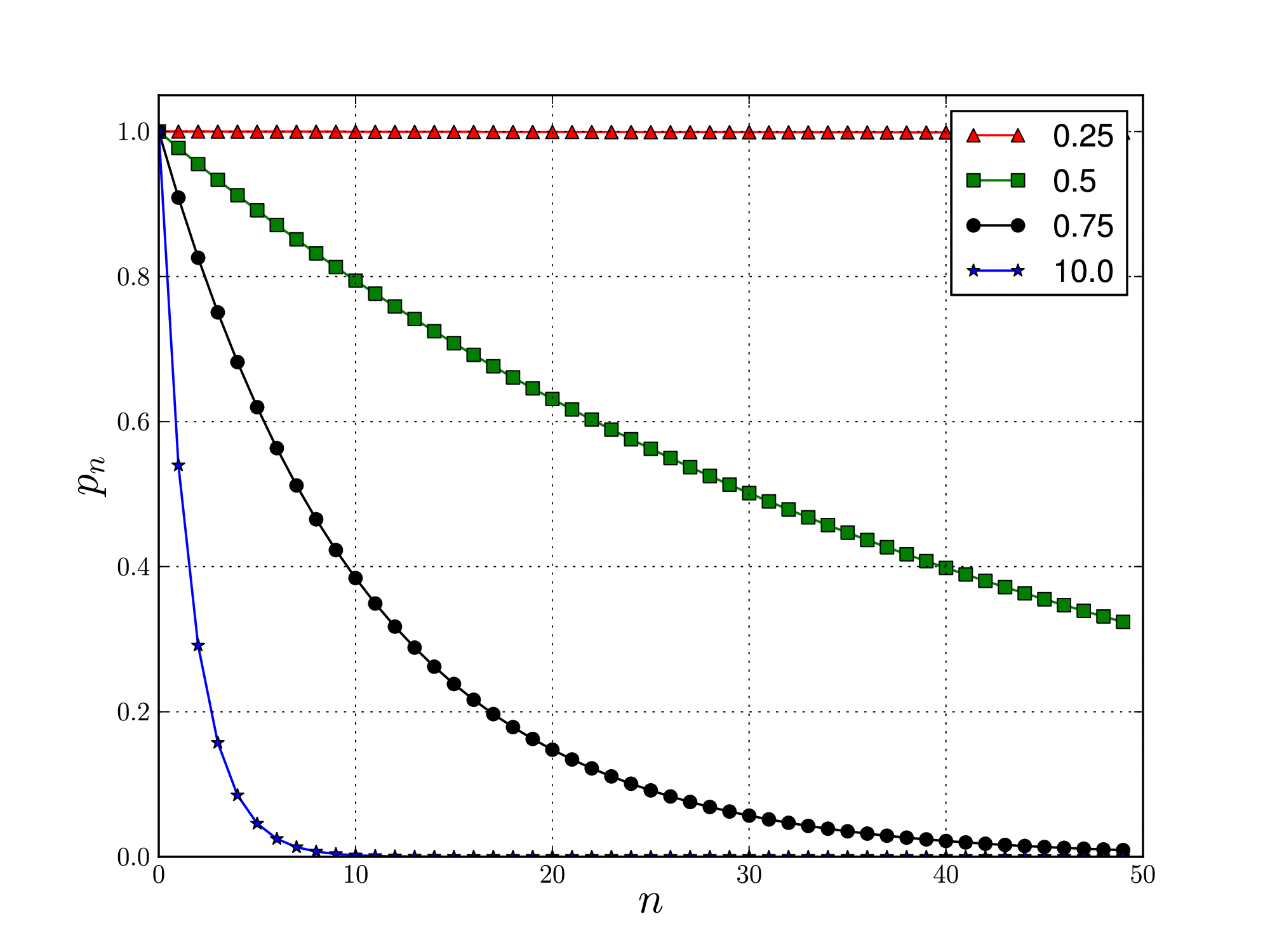}
\caption{The probability $p_n$ for values $ \sigma=0.25,\ldots10.0$, the fast decaying curve implies least persistence.}
\end{center}
\end{figure}

\section*{Acknowledgements}
I wish to thank Jim Chacko for the useful discussions and suggestions.

\section{Conclusion}
We have studied a model of CGRW in which correlation between steps has been introduced by a probability density function of the individual step. Some limiting cases of CGRW in one dimension were studied in greater detail where we were able to find the probability densities of displacement $S_n$. We note that the models we have studied in Sec. 3.1 and 3.2 are two particular cases of CGRW. It is only through experiments that one can get a clue of how to design an appropriate `strategy' function $\mu$.  In the model in Sec. 3.1 we studied the CGRW in one dimension in which the consecutive steps were correlated only in their directions. We found that this walk is persistent. Next we investigated a model in which the correlation depends also on the length of steps. Here we found that the random walk trajectories consists of clusters of walk and long jumps. This showed an oscillating behavior in the probability density. It was found that this walk did not show directional persistence. It would also be interesting to see the effect of time dependent memory in these walks. Further investigations can be carried out to see whether new interesting features arise in higher dimensions. We were able to identify parameters which could be associated with the 'strategy' and 'memory' of the random walker. In one of the cases we saw that correlation brings directional persistence in random walk but it was also shown that directional persistence may not be seen for any arbitrary `strategy' used by the random walker with a finite `memory'. This model has a potential application to a wide range of animal dispersal.

\end{document}